\documentclass[
aps,                    
prd,                    
showpacs,               
superscriptaddress,    
nofootinbib,            
twocolumn,             %
showkeys,               %
preprintnumbers,        %
amsmath,               %
amssymb,               %
floatfix]               
{revtex4-1}               

\usepackage{graphicx}
\usepackage{bm}
\usepackage{multirow}
\usepackage{array}
\usepackage{color}
\usepackage{ulem}
\usepackage{slashed}

\newcommand{\TA}{\ensuremath{\Tilde{A}}}
\newcommand{\Ta}[1]{\ensuremath{\Tilde{\alpha}_{#1}}}
\newcommand{\iTa}[1]{\ensuremath{\imath\Tilde{\alpha}_{#1}}}
\newcommand{\dTa}[1]{\ensuremath{\delta\Tilde{\alpha}_{#1}}}
\newcommand{\idTa}[1]{\ensuremath{\imath\delta\Tilde{\alpha}_{#1}}}

\newcommand{\Tb}{\ensuremath{\Tilde{\beta}}}
\newcommand{\iTb}{\ensuremath{\imath\Tilde{\beta}}}

\newcommand{\Hee}{\ensuremath{H_{ee}}}
\newcommand{\Hem}{\ensuremath{H_{e\mu}}}
\newcommand{\Hme}{\ensuremath{H_{\mu e}}}
\newcommand{\Hmm}{\ensuremath{H_{\mu\mu}}}

\newcommand{\Tk}[1]{\ensuremath{\Tilde{k}_{#1}}}
\newcommand{\TK}{\ensuremath{\Tilde{K}}}
\newcommand{\dTk}[1]{\ensuremath{\delta\Tilde{k}_{#1}}}

\newcommand{\dQ}[1]{\ensuremath{\delta Q_{#1}}}

\newcommand{\Tt}[1]{\ensuremath{\Tilde{\theta}_{#1}}}

\newcommand{\TU}{\ensuremath{\Tilde{U}}}

\begin{document}
\topmargin0.1in
\bibliographystyle{apsrev}
\title{The neutrino-neutrino interaction effects in supernovae: the point of view from the 'matter' basis}

\author{S\'ebastien Galais}
\email{galais@ipno.in2p3.fr}
\affiliation{Institut de Physique Nucl\'eaire, F-91406 Orsay cedex, CNRS/IN2P3 and University of Paris-XI, France}

\author{James Kneller} 
\email{jpknelle@ncsu.edu}
\affiliation{Department of Physics, North Carolina State University, Raleigh, NC 27695, USA}

\author{Cristina Volpe}
\email{volpe@ipno.in2p3.fr}
\affiliation{Institut de Physique Nucl\'eaire, F-91406 Orsay cedex, CNRS/IN2P3 and University of Paris-XI, France}

\date{\today}

\pacs{14.60.Pq, 97.60.Bw}

\begin{abstract}
We consider the Hamiltonian for neutrino oscillations in matter 
in the case of arbitrary potentials including off-diagonal complex terms.
We derive the expressions for the corresponding Hamiltonian in the basis of the instantaneous eigenstates in matter,
in terms of quantities one can derive from the flavor basis Hamiltonian and its derivative, for an arbitrary number of neutrino flavors.
We make our expressions explicit for the two-neutrino flavor case and apply our results to the neutrino propagation in core-collapse supernovae where the Hamiltonian includes both coupling to matter and to neutrinos. 
We show that the neutrino flavour evolution depends on the mixing matrix derivatives involving not only the derivative of the matter mixing angles but also of the phases. In particular, we point out the important role of the phase derivatives, that appear due to the neutrino-neutrino interaction, and show how it can cause an oscillating degeneracy between the diagonal elements of the Hamiltonian in the basis of the eigenstates in matter. Our results also reveal, that the end of the synchronization regime is due to a rapid increase of the phase derivative, and identify the condition to be fulfilled for the onset of bipolar oscillations involving both the off-diagonal neutrino-neutrino interaction contributions and the vacuum terms. 
\end{abstract}

\maketitle


\section{Introduction} \label{sec:intro}
\noindent 
Successful core-collapse supernova explosions seem to be at reach in the next decade. Currently, sophisticated two and three dimensional simulations comprise convection, hydrodynamic instabilities (in particular the SASI mode), realistic nuclear networks and neutrino transport. It appears that the explosion mechanism is being delineated for a number of progenitor masses \cite{Marek:2007gr,Bruenn:2010af}.
   
Since the discovery of neutrino oscillations, neutrino experiments have determined most of the parameters of the Maki-Nakagawa-Sakata-Pontecorvo matrix \cite{Maki:1962mu}, relating the flavor to the mass basis. The focus of next generation experiments is to measure the third mixing angle, the (Majorana and Dirac) CP violating phases,
the absolute mass scale and hierarchy, the Dirac versus Majorana neutrino nature \cite{Giunti:2007ry}. 
The experimental progress has numerous implications, e.g. for the neutrino flavor conversion in media. It is now clear that the solar neutrino deficit  is due to Mikheev-Smirnov-Wolfenstein (MSW) effect 
\cite{Wolfenstein1977,M&S1986} - the resonant conversion produced by the neutrino interactions while travelling through matter. 

The investigations of the neutrino flavor conversion in core-collapse supernovae have revealed an unexpected complexity compared to the solar case.
This is due to the realization of calculations including the neutrino-neutrino interaction 
\cite{Pantaleone:Gamma1292eq,Samuel:1993uw}, using matter density profiles with shocks \cite{Schirato:2002tg,Takahashi:2002yj,Fogli:2003dw,Tomas:2004gr,Choubey:2006aq,Kneller:2007kg,Gava:2009pj} and having turbulence \cite{1990PhRvD..42.3908S,Loreti:1995ae,Fogli:2006xy,Friedland:2006ta,Kneller:2010ky,Kneller:2010sc}. 
Understanding how the neutrino flavour conversion in supernovae is modified by the neutrino-neutrino interaction is a key theoretical and phenomenological issue. This question is at present being intensively investigated. 
It is now clear that three flavour conversion regimes exist. They consist first in a collective synchronization of the neutrinos with no flavor conversion \cite{Duan:2005cp,Pastor:2001iu}, then in the occurrence of "bipolar" flavour oscillations \cite{Duan:2005cp,Hannestad:2006nj} and finally in a complete swap of the (anti) neutrino spectra above a critical energy (the spectral split) \cite{Duan:2005cp,Duan:2006an,Raffelt:2007cb} (for a review see \cite{Duan:2009cd,Duan:2010bg}). 

Several works have investigated the conditions for the neutrino-neutrino effects to be triggered and for the onset of the bipolar oscillations. Studies are available both in simplified models \cite{Hannestad:2006nj} and in more sophisticated three-flavour and multi-angle simulations \cite{Dasgupta:2010cd,Duan:2010bf}.
Using the analogy with a pendulum, in \cite{Hannestad:2006nj} it has been pointed out that, contrarily e.g. to the well known MSW effect, collective flavour conversion effects occur for any value of the third neutrino mixing angle, while such a parameter needs to be stricly non-zero.  Besides 
the authors have shown
  that bipolar oscillations start in inverted neutrino mass hierarchy, because the flavor polarization vector is in an unstable position. However, it has recently emerged, that such instabilities are a more general behaviour and can be present for any neutrino mass hierarchy, depending on the primary neutrino fluxes \cite{Dasgupta:2009mg, Fogli:2009rd,Dasgupta:2010cd}. They can induce not only single but also multiple spectral splits. Another step in identifying the conditions for the bipolar oscillations to start has been done in \cite{Duan:2010bf} where an heuristic condition involving the vacuum and the neutrino-neutrino interaction terms is identified.

Most of the available calculations of neutrino propagation in supernovae including the neutrino-neutrino interaction assume the so-called "single-angle" approximation where the flavour evolution history is trajectory independent; while "multi-angle" calculations consider the flavour history along different trajectories and different interaction angles are considered. While the former has been shown to catch well qualitatively and quantitavively many features of the multi-angle calculations \cite{Fogli:2007bk}, the two calculations reveal differences that can be important. In particular in multi-angle calculations collective effects present flavour decoherence, as discussed e.g. in \cite{EstebanPretel:2007ec}; while, when the matter density exceeds (is comparable) the neutrino density, collective effects can be strongly suppressed (or be affected by multi-angle decoherence)\cite{EstebanPretel:2008ni}. 
In Ref.\cite{Duan:2010bf} the authors have also shown that the location where bipolar oscillations start varies if a single-angle or a multi-angle calculation is performed. The implications of different onset locations of the bipolar oscillations on the r-process has been further investigated in \cite{Duan:2010af} where it was shown that multi-angle versus single-angle calculations can produce different r-process abundances. Note that the importance of the neutrino-neutrino interaction for the nucleosynthesis of heavy elements was identified in an early work \cite{Balantekin:2004ug}. In the long run it is clear that accurate theoretical predictions will be based both upon full multi-angle calculations and also consider a non-spherical geometry for the neutrinosphere, which can be important in astrophysical environments such as accretion disks around black holes and coalescing neutron stars \cite{Dasgupta:2008cu}.

 In addition to the sensitivity to the details of the neutrino spectra at the neutrinosphere and the neutrino hierarchy, the results in presence of the neutrino-neutrino interaction can also be sensitive to the Dirac CP violating phase, due to loop corrections or to physics beyond the Standard Model, as demonstrated in \cite{Balantekin:2007es,Gava:2008rp}. While first studied in \cite{Akhmedov:2002zj}, the existence and conditions for possible CP violating effects in supernovae has been established in \cite{Balantekin:2007es} and \cite{Gava:2008rp}, in presence of the neutrino-neutrino interaction, and a first quantification of the CP violating Dirac phase impact on the neutrino fluxes has been performed. The analytical results  in \cite{Balantekin:2007es} are now independently confirmed in \cite{Kneller:2009vd}. Note that Ref. \cite{Gava:2010kz} has explored the impact of a non-zero Dirac phase on the neutrino degeneracy parameter at the Big-Bang nucleosynthesis epoch, extending the work done in \cite{Balantekin:2007es,Gava:2008rp}.
In conclusion, at present, many features of the neutrino-neutrino interaction effects have been understood and some of the phenomenological implications explored, but many questions remain. In particular, work is still needed to fully unravel the physical mechanisms underlying the neutrino flavor conversion in presence of such contributions, and also their interplay with the unknown neutrino properties.

The goal of the present work is to try to gather further insight
in the neutrino-neutrino effects using the basis of the instantaneous eigenstates in matter,
instead of the flavour one.
To this aim we follow the neutrino evolution through matter and present general expressions for the diagonal and off-diagonal terms of the Hamiltonian describing neutrino evolution in the basis of the eigenstates in matter, for an arbitrary number of neutrino families and for the case of an arbitrary Hamiltonian - in the flavor basis - having in particular complex off-diagonal contributions. We define the corresponding generalized non-adiabaticity parameters. This part extends a previous work \cite{Kneller:2009vd}. While the relations we find are general, we apply our results to the case of neutrino evolution in core-collapse supernovae, and give explicit expressions in the two neutrino flavor case. We then point out for the first time the important role of the matter mixing matrix Dirac phase, engendered by the presence of the neutrino-neutrino interaction contribution to the Hamiltonian. Numerical calculations are provided of the diagonal and off-diagonal terms of the Hamiltonian in the basis of the eigenstates in matter. 
We show that the start of the bipolar oscillations is associated with a rapid growth of the phase derivative and identify an analytical condition that involves the neutrino-neutrino interaction and vacuum terms. 

Our paper is organized as follows. The formalism and our generalized expressions for the non-adiabaticity parameter in the basis of the eigenstates in matter is presented in Section II, followed by a focus upon the two flavor case as an example. In Section III we apply our results to the case of neutrino propagation in supernova where the Hamiltonian is composed of the coupling to 
matter and to neutrinos and present numerical results in two flavors for the diagonal and off-diagonal entries of the Hamiltonian in the basis of eigenstates in presence of matter. Section IV is a conclusion. Appendix A shows that the Majorana phases in matter do not change our conclusions. Appendix B provides the expressions for the derivative of the neutrino-neutrino Hamiltonian in the multiangle case.

\section{The Formalism}
\noindent
In this section we derive general expressions 
for the Hamiltonian, in the basis of the instantaneous eigenstates in matter, describing
neutrino propagation in an environment when the Hamiltonian in the flavour basis contains complex potentials. We will call from now on such a basis the 'matter' basis. We do not make 
any assumption about the Hamiltonian entries, but only that the Hamiltonian and its derivative can be computed. To make expressions more explicit we consider
neutrino propagation in the framework of core-collapse supernovae.
\subsection{Neutrino Evolution in the Flavor Basis} 
\noindent
The neutrino evolution is determined by the Schr\"odinger equation ($\hbar=c=1$) :
\begin{eqnarray}\label{e:1}
i \frac{d\psi^{(f)}}{dt} & = & H^{(f)}  \psi^{(f)}
\end{eqnarray}
where $ \;\psi^{(f)}$ are the neutrino amplitudes for a neutrino to be in a given flavour state, for an arbitrary number $N$ of neutrino families,
$H^{(f)}$ the Hamiltonian in the flavour basis. The latter is composed of multiple terms,
\begin{equation}\label{e:1}
H^{(f)} = U\,K\,U^{\dag}+ H_{mat}^{(f)}+ H^{(f)}_{\nu\nu} +\ldots
\end{equation}
 namely the rotated vacuum Hamiltonian $U\,K\,U^{\dag}$, 
the potentials $H_{mat}^{(f)}$ due to the coupling of neutrinos with matter, which are diagonal in the flavor basis, and the neutrino-neutrino 
interaction term $H^{(f)}_{\nu\nu}$ which is not, in general, diagonal in the flavor basis. 
The Hamiltonian $K$ is given by 
$K = diag(k_1, k_2, k_3..., k_N) $ with $k_i$ being the neutrino energy eigenvalues in the mass basis.

In Eq.(\ref{e:1}) $U$ is the unitary matrix relating the flavor and
the mass basis, i.e. $| \nu_{\alpha} \rangle = \sum_{i=1,N} U_{\alpha i} | \nu_{i} \rangle$. 
For $N=3$ the $U$ matrix is the well known Maki-Nakagawa-Sakata-Pontecorvo matrix \cite{Nakamura:2010zzi}.
Obviously the elements of $U$ are related to each other via the requirement of unitarity. This allows us to express some 
elements in terms of others after specifying the phase of the determinant, and restrict the magnitude of the 
remaining independent elements. Thus the $N \times N$ mixing matrix $U$ can be parametrized in terms of $N(N-1)/2$ mixing angles and $N(N+1)/2$ phases, out of which $N$ are of Majorana while $N-1$ are of Dirac type\footnote{The assignment of the phases to Dirac or Majorana type is not unique: different parameterizations will change the assignment.}. 
This leaves $(N-1)(N-2)/2$ leftover, CP phases. The dependence upon the Majorana phases, which we label as $\alpha_{i}$ can be factored in the mixing matrix: i.e. we write $U$ as $U = \slashed{U}\,A$ with $\slashed{U}$ independent of the $\alpha$'s and the matrix $A$ defined to be $A=diag(\exp(-i\alpha_{1}),  \exp(-i \alpha_{2}), ...,  \exp(-i \alpha_{N}))$ \cite{Nakamura:2010zzi}.
The $N$ Majorana phases and the matrix $A$ have no role in neutrino oscillations \cite{1987NuPhB.282..589L} because all observables are given by the squared modulus of a matrix element and the Majorana phases only ever enter through the phases of the elements.

It is well known that the Dirac phases can be absorbed by redefining the charged fermion fields in the standard model Lagrangian. This possibility indicates that their absolute values cannot affect observables, however, once this is done, we have removed this degree of freedom. While this rephasing can certainly be performed when neutrinos evolve in vacuum, in matter this procedure might not be approriate, in particular when the neutrino propagation Hamiltonian comprises off-diagonal complex terms, as is the case in presence of the contribution coming from the neutrino-neutrino interaction. Indeed, we will show that such phases can play an important role.

If we restrict to $N=3$ families and consider that neutrinos interact via the standard weak interaction with ordinary matter composed of electrons, protons and neutrons, the contribution to $H^{(f)}$ coming from $H_{mat}^{(f)}$ is diagonal in the flavor basis 
$V^{(f)}(\mathbf{r})=diag(V_{e}(\mathbf{r}),0,0)$ with $V_{e}(\mathbf{r}) =\sqrt{2}\,G_{F}\,n_e(\mathbf{r})$ and
$n_e(\mathbf{r})$ the electron density, $\mathbf{r}$ being the distance in the supernova. 
The contribution coming from scattering on neutrons can be
substracted. In fact it is possible to remove the trace of this Hamiltonian because we are free to add or subtract 
an arbitrary multiple of the unit matrix (including a term that is a function of position), 
since the only effect of such a term is to introduce an overall phase. The third term in $H^{(f)}$ correspond to the 
neutrino-neutrino interaction Hamiltonian \cite{Duan:2006an}
\begin{eqnarray}\label{e:1tris}
H^{(f)}_{\nu\nu} &= &\sqrt{2} G_F \sum_{\alpha}\int \rho_{\nu_{\underline{\alpha}}}({\bf q'})(1- \hat{\bf{q}} \cdot \hat{\bf{q}}') dn_{\nu_{\underline{\alpha}}} dq'\\ \nonumber
 && - \sqrt{2} G_F \sum_{\alpha}\int \bar{\rho}_{\bar{\nu}_{\underline{\alpha}}}({\bf q'})(1- \hat{\bf{q}} \cdot \hat{\bf{q}}') dn_{\bar{\nu}_{\underline{\alpha}}} dq'
\end{eqnarray}
where $d{n}_{\nu_{\underline{\alpha}}}$ ($d{n}_{\bar{\nu}_{\underline{\alpha}}}$) are the differential neutrino (anti-neutrino) number densities and $\rho_{\nu_{\underline{\alpha}}}$ ($\bar{\rho}_{\bar{\nu}_{\underline{\alpha}}}$) is the density matrix for neutrinos (anti-neutrinos), enconding neutrino flavour conversion, 
whose expression is e.g. for neutrinos
\begin{eqnarray}
\label{cosmo1}
\rho_{{\nu}_{\alpha}}\equiv\left(\begin{array}{ccc}
|\nu_e|^2  &  \nu_e\nu_{\mu}^*   & \nu_e\nu_{\tau}^* \\
  \nu_e^*\nu_{\mu}    &  |\nu_{\mu}|^2 &  \nu_{\mu}\nu_{\tau}^*  \\
   \nu_e^*\nu_{\tau}    &  \nu_{\mu}^*\nu_{\tau}  & |\nu_{\tau}|^2 
\end{array} \right)
\end{eqnarray}
and similalrly for $\bar{\rho}_{\bar{\nu}_{\underline{\alpha}}}$.
In Eq.(\ref{e:1tris}) the neutrino-neutrino interaction term is built from the contributions coming from neutrinos and anti-neutrinos having different momenta $\bf{q}, \bf{q'}$, $\hat{\bf{q}}= \bf{q}/|\bf{q}|$ with and born with the flavour $\alpha$ at the neutrinosphere\footnote{The fact that the neutrino flavour content might have changed is indicated with a flavour underlined.}.  The fact that their flavour might have changed up to the interaction point is encoded in the neutrino density matrix.  

\subsection{Neutrino Evolution in the 'Matter' Basis}
\noindent
We now consider the basis that instantenously diagonalizes the Hamiltonian $ H^{(f)} $ Eq.(\ref{e:1})
\begin{equation}\label{e:3}
\TU^{\dag}\,H^{(f)}\,\TU = \Tilde{K} =\left(\begin{array}{lll} \Tk{1} & 0
& \ldots \\ 0 & \Tk{2} & \ldots \\ \vdots & \vdots & \ddots \end{array}\right)
\end{equation}
where $\Tk{i}, i=1,N$ are the corresponding energy eigenvalues in matter.  For the two and three flavor cases, the 
expressions for the eigenvalues are well known \cite{Betal80,Ohlsson:1999xb} and one may also find an algebraic expression for $N = 4$. For $N\geq 5$ the Abel-Ruffini theorem indicates that no general algebraic formula exists.
Once the eigenvalues are found, one can determine the unitary
transformation $\TU$ relating the flavour basis and the 'matter' basis.
Note that,  to distinguish quantities evaluated in vacuum from those calculated in matter, in the following we will use symbols with tilde, to signify that e.g. the energy eigenvalues as well as the parameters (phases and angles) of the $\Tilde{U}$ matrix are those evaluated in presence of matter.
Like $U$, the matrix $\TU$ also has $N$ Majorana phases $\Ta{i}$ that we can factor out from $\TU$ so that $\TU=\slashed{\TU}\TA$ with $\TA=diag( \exp(-i\Tilde{\alpha}_{1}),  \exp(-i\Tilde{\alpha}_{2}),  , ...,  \exp(-i \Tilde{\alpha}_{N}))$.    
If the eigenvalues are non-degenerate then we can derive the identity that the elements of 
$\TU$ satisfy: 
\begin{equation}\label{e:4}
\TU_{\alpha i}^{\ast}\,\TU_{\beta i} = \frac{C^{(i)}_{\alpha\beta}}{\sum_{\gamma}C^{(i)}_{\gamma\gamma}}  
\end{equation}
where $C^{(i)}$ is the co-factor matrix of $H^{(f)}-\Tk{i}$ and so
\begin{equation}\label{e:Ubi2}
|\TU_{\beta i}|^{2}=\frac{C^{(i)}_{\beta\beta}}{\sum_{\gamma}C^{(i)}_{\gamma\gamma}}.
\end{equation} 
Note that both equations are independent of the $\Tilde{\alpha}$'s. 
The denominator of Eqs.(\ref{e:4}-\ref{e:Ubi2}) is the trace of the cofactor matrix $H^{(f)}-\Tk{i}$ which is basis independent, so that 
\begin{equation}\label{e:TraceC}
\sum_{\gamma}C^{(i)}_{\gamma\gamma}=\sum_{j}C^{(i)}_{jj}(\Tilde{K}-\Tk{i}) = \prod_{j\neq i}(\Tk{j}-\Tk{i})
\end{equation} 
In practice we use Eq.(\ref{e:4}) to relate the elements in a column to one specific element $\TU_{\beta i}$, then use Eq.(\ref{e:Ubi2}) to evaluate $\TU_{\beta i}$\footnote{Note that since the $N$ Majorana phases {\bf $\Ta{i}$} do not appear in Eqs.(\ref{e:4}-\ref{e:Ubi2}), we have no way to determine them and can make any choice 
including functions of the position (see Appendix A).}

Once $\TU$ is found, a change of basis can be made using $\TU$, to give the Schr\"odinger equation in the 'matter' basis
\begin{subequations}\label{e:7}
\begin{eqnarray}
i \frac{d\tilde{\psi}}{dt} & = & \left( \Tilde{K} -i
\TU^{\dag}\frac{d\TU}{dt} \right)\;\tilde{\psi}
\\
& = & \tilde{H}\;\tilde{\psi}
\end{eqnarray}
\end{subequations}
where $\tilde{H}$ is the Hamiltonian in the basis of the eigenstates in matter.
The term $\TU^{\dag}\,d\TU/dt$ appears because the matter eigenvalues are functions of
time, which requires that $\TU$ also be a function of time. In order to evaluate
$\tilde{H}$ we need to compute $\TU^{\dag}\,d\TU/dt$. 
The derivation
of the off-diagonal elements of $\TU^{\dag}\,d\TU/dt$ is straight-forward. 
In fact, if we differentiate the eigen-equation $H^{(f)} \TU =\TU\TK$ and multiply the result by $\TU^{\dag}$, we derive the 
result that
\begin{equation}\label{e:8}
\TU^{\dag}\,\frac{dH^{(f)}}{dt}\,\TU+\left[\TK,\TU^{\dag}\frac{d\TU}{dt}\right] =\frac{d\TK}{dt}
\end{equation}
The commutator vanishes for the diagonal elements of this equation, because $\TK$ is diagonal. Thus we find the result
that
\begin{equation}\label{e:dkidx}
\sum_{\alpha,\beta}\TU^{\star}_{\alpha i}\,\frac{dH^{(f)}_{\alpha\beta}}{dt}\,\TU_{\beta i} =\frac{d\Tk{i}}{dt}
\end{equation}
By considering the off-diagonal elements for this same expression, 
the following equation is deduced
\begin{equation}\label{e:10}
\left( \TU^{\dag}\,\frac{dH^{(f)}}{dt}\,\TU \right)_{ij} +\left[\TK,\TU^{\dag}\frac{d\TU}{dt}\right]_{ij}=0, \quad i\neq j
\end{equation}
Again, since $\TK$ is diagonal, the off-diagonal elements of the commutator have a compact form :
\begin{equation}\label{e:11}
\left(\TU^{\dag}\frac{d\TU}{dt}\right)_{ij}=-\frac{1}{\dTk{ij}}\,\left(\TU^{\dag}\,\frac{dH^{(f)}}{dt}\,\TU \right)_{ij}\\
\end{equation} 
where $\dTk{ij}$ is $\dTk{ij}=\Tk{i}-\Tk{j}$. 

The remaining missing pieces of $\tilde{H}$ 
are the diagonal terms of $\TU^{\dag}\,d\TU/dt$.
Because of the Majorana phase ambiguity mentioned previously there is no unique expression for these terms but, by using Eq.(\ref{e:4}),
we can eliminate the ambiguity from all elements of a column of $\TU$ bar one, $\TU_{\beta i}$.
Using the identity that $\TU^{\dag}\,d\TU/dt=-d\TU^{\dag}/dt\,\TU$ we derive that 
\begin{eqnarray}\label{e:12}
\left(\TU^{\dag}\frac{d\TU}{dt}\right)_{ii}&=&\frac{1}{2|\TU_{\beta i}|^{2}}\sum_{\alpha\neq\beta}\left[ \TU_{\alpha i}^{\ast}\,\TU_{\beta i}\frac{d}{dt}\left(\TU_{\beta i}^{\ast}\,\TU_{\alpha i}\right) \right.\nonumber \\
&&\left.\qquad\qquad\quad-\TU_{\beta i}^{\ast}\,\TU_{\alpha i}  \frac{d}{dt}\left(\TU_{\alpha i}^{\ast}\,\TU_{\beta i}\right) \right] \nonumber \\
&& + \frac{1}{2|\TU_{\beta i}|^{2}}\,\left( \TU_{\beta i}^{\ast}\frac{d\TU_{\beta i}}{dt} -\TU_{\beta i}\frac{d\TU_{\beta i}^{\ast}}{dt} \right)
\end{eqnarray} 
The first term of Eq.(\ref{e:12}) is well defined because it is independent of the $\Tilde{\alpha}$'s; while the Majorana phase ambiguity is entirely contained 
in the second term. Introducing the variable $Q_i$ to represent the first term 
in Eq.(\ref{e:12}), and using Eq.(\ref{e:4}), we find 
\begin{equation}\label{e:Qi}
i Q_{i}=\Xi\,\sum_{\alpha\neq\beta}\left( C^{(i)}_{\alpha\beta}\frac{dC^{(i)}_{\beta\alpha}}{dt} -C^{(i)}_{\beta\alpha}\frac{dC^{(i)}_{\alpha\beta}}{dt} \right)
\end{equation} 
with $\Xi=(2|\TU_{\beta i}|^{2}\,\left(Tr(C^{(i)})\right)^{2})^{-1}$. This leaves the second term in Eq.(\ref{e:12}). 
If the phase of $\TU_{\beta i}$ is chosen to be solely the Majorana phase i.e $arg(\TU_{\beta i}) = -\alpha_{i}$ - since there are $N-1$ Dirac phases $\beta$ will be fixed to the electron flavour for all $i$ - then 
\begin{equation}
\frac{1}{2|\TU_{\beta i}|^{2}}\,\left( \TU_{\beta i}^{\ast}\frac{d\TU_{\beta i}}{dt} -\TU_{\beta i}\frac{d\TU_{\beta i}^{\ast}}{dt} \right) = -i\frac{d\alpha_{i}}{dt}.
\end{equation}
Thus we obtain the result that the diagonal elements of $\TU^{\dag}\,d\TU/dt$ are 
\begin{equation}
\left(\TU^{\dag}\frac{d\TU}{dt}\right)_{ii} = i Q_{i} -i\,\frac{d\alpha_{i}}{dt}
\end{equation}
Inserting this equation into $\tilde{H}$ we obtain our final result that it can be written as 
\begin{widetext}
\begin{equation}\label{e:14}\renewcommand{\arraystretch}{1.5}
\tilde{H} =
\left(\begin{array}{cccc} 
\Tk{1}+Q_{1}-i\frac{d\Ta{1}}{dt} & i\,\frac{(\dTk{12}+\dQ{12})}{2\,\pi}\,\Gamma_{12} & i\,\frac{(\dTk{13}+\dQ{13})}{2\,\pi}\,\Gamma_{13}&\ldots \\ 
-i\frac{(\dTk{12}+\dQ{12})}{2\,\pi}\,\Gamma^{\ast}_{12} & \Tk{2}+Q_{2}-i\frac{d\Ta{2}}{dt} &i\,\frac{(\dTk{23}+\dQ{23})}{2\,\pi}\,\Gamma_{23}&\ldots \\ 
-i\,\frac{(\dTk{13}+\dQ{13})}{2\,\pi}\,\Gamma^{\ast}_{13} & -i\,\frac{(\dTk{23}+\dQ{23})}{2\,\pi}\,\Gamma^{\ast}_{23} & \Tk{3}+Q_{3}-i\frac{d\Ta{3}}{dt} &\ldots \\
\vdots & \vdots & \vdots & \ddots
\end{array}\right)
\end{equation}
\end{widetext}
with $\dQ{ij}$ given by $\dQ{ij}=Q_{i}-Q_{j}$
and we have introduced the 
$\Gamma_{ij}$ functions which
are the generalized non-adiabaticity parameters 
for neutrino oscillations 
with arbitrary potentials. These quantities are defined as 

\begin{eqnarray}\label{e:Gammaij}
\Gamma_{ij}  & = &  -\frac{2\,\pi e^{i\delta \alpha_{ij}}}{(\dTk{ij}+\dQ{ij})}\left(\TU^{\dag}\frac{d\TU}{dt}\right)_{ij} \\ \nonumber
& = & \frac{2\,\pi e^{i\delta \alpha_{ij}}}{\dTk{ij}\,(\dTk{ij}+\dQ{ij})}\left(\TU^{\dag}\,\frac{dH^{(f)}}{dt}\,\TU \right)_{ij}
\end{eqnarray}

\noindent
where we have substituted in, the results from Eq.(\ref{e:11}). 
The non-adiabaticity paramaters $\Gamma_{ij}$ have, in general, both real and imaginary components whose origins are from different components in the 
flavor-basis Hamiltonian. Let us investigate this connection further by considering the case of two-neutrino flavors. 
While many aspects are familiar from the MSW problem,
new components arise when we consider more general Hamiltonians.


\subsection{The application to two-neutrino flavors} 
\label{sec:2flavour}
\noindent
Let us consider
the general parametrization of the mixing matrix in matter for two neutrino flavours 
\begin{equation}\label{e:U2}
\TU = 
\left(\begin{array}{ll} 1 & 0  \\ 0 & e^{\iTb} \end{array}\right)
\left(\begin{array}{rr} \cos\Tt{} & \sin\Tt{}  \\ -\sin\Tt{} & \cos\Tt{} \end{array}\right)
\left(\begin{array}{ll} e^{-\iTa{1}} & 0 \\ 0 & e^{-\iTa{2}}\end{array}\right)
\end{equation} 
and note that the phase $\tilde{\beta}$ is the Dirac phase in matter. With this parametrization the elements of $H^{(f)}$ are given by\footnote{Note how the Majorana phases do not appear in these expressions.}
\begin{eqnarray}\label{e:Hf}
H^{(f)} & = & \left(\begin{array}{ll}  \Hee &   \Hem \\  
\Hme & \Hmm \end{array}\right) \\ \nonumber
 &= &\left(\begin{array}{l r}  \Tk{1}\,\cos^{2}\Tt{} +\Tk{2}\,\sin^{2}\Tt{} &  - \dTk{12}\,e^{-i\Tb}\,\cos\Tt{}\,\sin\Tt{}  \\  -\dTk{12}\,e^{i\Tb}\,\cos\Tt{}\,\sin\Tt{} &   \Tk{1}\,\sin^{2}\Tt{} +\Tk{2}\,\cos^{2}\Tt{}\end{array}\right)
\end{eqnarray}

From Eqs.(\ref{e:Hf}) it is very simple to derive several different relationships between the matter angle $\Tt{}$ and
the elements of the flavor basis Hamiltonian e.g.
\begin{equation}\label{e:theta}
\sin^2\Tt{}=\frac{\Hee- \Tk{1}}{ \Tk{2} - \Tk{1} }
\end{equation} 
The expression for $\cos^2\Tt{}$ is very similar to equation (\ref{e:theta}) but with $\Hmm$ in place of $\Hee$. For the Dirac phase 
we derive that  
\begin{equation}\label{e:22}
\tan\tilde{\beta}=i\left( \dfrac{\Hem-\Hme}{\Hem+\Hme}\right)  = -\dfrac{\mathfrak{I}(\Hem)}{\mathfrak{R}(\Hem)} 
\end{equation} 
From this equation we immediately see that in the case when the off-diagonal elements of $H^{(f)}$ are independent of position then $\Tb$ is simply the same as its vacuum value.

By differentiating $\TU$ and multiplying by $\TU^{\dag}$ one finds\footnote{We have omitted the derivatives of the Majorana phases.}
\begin{eqnarray}\label{e:UdagdUdx}
\TU^{\dagger} \frac{d\TU}{dt} & = &
i  \left(\begin{array}{c c} \sin^{2}\Tt{} & -\frac{\sin2\Tt{}}{2}\,e^{\idTa{12}}  \\ -\frac{\sin2\Tt{}}{2}\,e^{-\idTa{12}} & \cos^{2}\Tt{} \end{array}\right)\;\frac{d\Tb}{dt} \nonumber\\
&&+ \left(\begin{array}{c c} 0 & e^{\idTa{12}}  \\ -e^{-\idTa{12}} & 0 \end{array}\right)\;\frac{d\Tt{}}{dt} 
\end{eqnarray}
so that, following Eq.(\ref{e:14}), the full Hamiltonian 
\begin{equation}\label{e:33}\renewcommand{\arraystretch}{2}
\tilde{H} = 
\left(\begin{array}{c c} \Tk{1} + Q_{1}  &i\frac{(\dTk{12}+\dQ{12})}{2\pi}\Gamma_{12}  \\
                        -i\frac{(\dTk{12}+\dQ{12})}{2\pi}\Gamma^{\ast}_{12} & \Tk{2} + Q_{2}   \end{array}\right)
\end{equation}
explicitly reads\footnote{From now on we indicate differentiation by $\dot{f}=df/dt$ for compactness.}
\begin{equation}\label{e:18}\renewcommand{\arraystretch}{1.5}
\tilde{H}  = 
  \left(\begin{array}{c c} \tilde{k}_1 +\dot{\tilde{\beta}}\sin^2\Tt{} & -e^{\idTa{12}}(\dot{\tilde{\beta}}\,\frac{\sin2\Tt{}}{2} +i\,\dot{\tilde{\theta}})   \\ - e^{-\idTa{12}}(\dot{\tilde{\beta}}\,\frac{\sin2\Tt{}}{2} - i\,\dot{\tilde{\theta}}) & \tilde{k}_2 +\dot{\tilde{\beta}}\cos^{2}\Tt{} \end{array}\right)\;
\end{equation} 
where $\dTa{12}=\Ta{1}-\Ta{2}$.
From the diagonal elements of $\tilde{H}$ we read off the $Q$'s to be
\begin{subequations}
\begin{eqnarray}\label{e:Q1Q2}
Q_{1}= \dot{\tilde{\beta}}\sin^2\Tt{} \\
Q_{2}= \dot{\tilde{\beta}}\cos^{2}\Tt{}
\end{eqnarray}
\end{subequations} 
and from the off-diagonal entries we derive 
that the generalised non-adiabaticity parameter is
\begin{equation}\label{e:Gamma12}
\Gamma_{12}=-\frac{2\pi\,e^{\idTa{12}}}{\dTk{12}+\dQ{12} }\left(\dot{\Tt{}}-i\,\frac{\sin2\Tt{}}{2}\,\dot{\Tb}\right)
\end{equation} 
and observe that it depends both on the derivative of the matter angle $\tilde{\theta}$, as in the MSW case, and on the derivative of the matter phase $\tilde{\beta}$. 
From Eq.(\ref{e:Gamma12}) one sees that the $\Gamma_{12}$ are defined up to the Majorana phases.
In order for the imaginary component of $\Gamma_{12}$ to be non-zero we require that the off-diagonal elements of $H^{(f)}$ rotate in the Argand plane. 
Neutrino-neutrino interaction gives exactly such a term because the flavor basis Hamiltonian includes contributions from the 
density matrices i.e. $H \sim \rho$.

We can then differentiate
Eqs.(\ref{e:Hf}) and eventually find
\begin{subequations}
\begin{eqnarray}
\dot{\Tk{}}_{1} & = & \cos^{2}\Tt{}\,\dot{H}_{ee} +\sin^{2}\Tt{}\,\dot{H}_{\mu\mu} \nonumber \\ 
&& \quad -\cos\Tt{}\,\sin\Tt{}\left(e^{i\Tb}\dot{H}_{e\mu}+e^{-i\Tb}\dot{H}_{\mu e}\right)  \\
\dot{\Tk{}}_{2} & = & \sin^{2}\Tt{}\,\dot{H}_{ee} +\cos^{2}\Tt{}\,\dot{H}_{\mu\mu} \nonumber \\
&& \quad +\cos\Tt{}\,\sin\Tt{}\left(e^{i\Tb}\dot{H}_{e\mu}+e^{-i\Tb}\dot{H}_{\mu e}\right)
\end{eqnarray}
\end{subequations} 
which is consistent with our expectations from Eq.(\ref{e:dkidx}). We can then differentiate the expressions for $\Tt{}$ and $\Tb$ and derive that  
\begin{equation}
\dot{\tilde{\theta}}=\frac{\left( \dot{H}_{ee}-\dot{H}_{\mu\mu}\right)\,\dTk{12} - (\Hee-\Hmm)\,\delta\dot{\tilde{k}}_{12}}{4\dTk{12}|\Hem|}
\end{equation} 
and
\begin{eqnarray}
\dot{\tilde{\theta}}&=& - \frac{\sin2\Tt{}(\dot{H}_{ee} - \dot{H}_{\mu\mu})}{2\dTk{12}} -\frac{\cos2\Tt{}}{2\dTk{12}}\left(e^{i\Tb}\dot{H}_{e\mu}+e^{-i\Tb}\dot{H}_{\mu e}\right)  \nonumber \\ 
&&\label{e:23b} 
\end{eqnarray}
The first term in Eq.(\ref{e:23b}) will be familiar to many readers because it is the MSW term, the second term is new.

From the two terms on the {\it r.h.s.} of Eq.(\ref{e:22}), two expressions for $\dot{\tilde{\beta}}$ can be obtained :
\begin{subequations}
\begin{eqnarray}
\dot{\tilde{\beta}} & = & \frac{-i}{\dTk{12}\sin2\Tt{}}\left(e^{i\Tb}\dot{H}_{e\mu} -e^{-i\Tb}\dot{H}_{\mu e}\right) \\ 
\dot{\tilde{\beta}}&=&\dfrac{\mathfrak{I}(\Hem)\mathfrak{R}(\dot{H}_{e\mu})- \mathfrak{R}(\Hem)\mathfrak{I}(\dot{H}_{e\mu}) }{|\Hem|^2}. \label{e:23} 
\end{eqnarray}
\end{subequations} 
If one substitutes into Eq.(\ref{e:Gamma12}) our results for $\dot{\Tt{}}$ and $\dot{\Tb}$, one gets
\begin{eqnarray}
\Gamma_{12}&=&\frac{2\pi\,e^{\idTa{12}}}{\dTk{12}(\dTk{12}+\dQ{12})}\left( \cos\Tt{}\sin\Tt{}\left(\dot{H}_{ee}-\dot{H}_{\mu\mu}\right) \right. \nonumber \\
&& \quad \left. +\dot{H}_{e\mu}\,\cos^{2}\Tt{}\,e^{i\beta} -\dot{H}_{\mu e}\,\sin^{2}\Tt{}\,e^{-i\beta} \right)
\end{eqnarray} 
The reader may verify that this is exactly the expression one would have obtained directly from equation (\ref{e:Gammaij}).

\section{The application to the neutrino-neutrino interaction problem}
\subsection{The phase derivative and the onset of the synchronization and bipolar regimes}
\noindent
Let us now apply the formalism and the results of the previous section to the Hamiltonian Eqs.(\ref{e:1}-\ref{e:1tris}) describing neutrino propagation in a core-collapse supernova with the neutrino-neutrino interaction contribution.
In the calculations we use a realistic matter density profile coming from supernova simulations. We treat the neutrino-neutrino contribution in the single-angle approximation. In Appendix B we provide equations for the derivative of the neutrino-neutrino interaction Hamiltonian in the multi-angle case. Note that from now on we replace time (t) with distance (x).
If we calculate the derivative of $H$ necessary to determine the diagonal and off-diagonal terms of the Hamiltonian in the 'matter' basis one has two contributions, namely
$ \dot{H}=\dot{H}_{mat}+\dot{H}_{\nu\nu}$, since the vacuum term does not contribute.

In the single-angle approximation, $H_{\nu\nu}$ Eq.(\ref{e:1tris}) reads
\begin{equation}\label{e:24}
H_{\nu\nu}^{s.a.}=F(x)^{s.a.}G(\rho)^{s.a.}
\end{equation} 
with the geometrical factor
 \begin{equation}\label{e:25}
F(x)^{s.a.}=\dfrac{\sqrt{2} G_F}{2 \pi R_{\nu}^2}\dfrac{1}{2}\left[1-\sqrt{1-\left(\dfrac{R_{\nu}}{x}\right)^2}\right]^2
\end{equation} 
where $R_{\nu}$ is the radius of the neutrinosphere. 
The non-linear contribution is given by
\begin{equation}\label{e:26}
G(\rho)^{s.a.}=\sum_{\alpha} \int (\rho_{\nu_{\underline{\alpha}}}(q')
L_{\nu_{\underline{\alpha}}}(q')
-\bar{\rho}_{\bar{\nu}_{\underline{\alpha}}}(q')
L_{\bar{\nu}_{\underline{\alpha}}}(q'))dq'
\end{equation} 
with $L_{\nu_{\underline{\alpha}}}$  the neutrino flux at the same location for a neutrino of a flavor $\alpha$. 
The corresponding derivative {\bf $\dot{H}_{\nu\nu}$} includes contributions  from both the derivative of the geometrical factor and the density matrices, i.e.
\begin{eqnarray}\label{e:27}
\dot{H}_{\nu\nu}^{s.a.}=\dot{F}(x)^{s.a.}G(\rho)^{s.a.} + F(x)^{s.a.}\dot{G}(\rho)^{s.a.}
\end{eqnarray}
where the calculation of $\dot{F}(x)$ is straightforward and the one for the non-linear term is obtained using of the Liouville Von-Neumann equation. This gives 
\begin{eqnarray}\label{e:28}
\dot{G}(\rho)^{s.a.} & = & - i \sum_{\alpha} \int ( [H,\rho_{\nu_{\underline{\alpha}}}(q')] L_{\nu_{\underline{\alpha}}} (q') \nonumber\\
&& \qquad\quad + [\bar{H},\rho_{\bar{\nu}_{\underline{\alpha}}}(q')]^* L_{\bar{\nu}_{\underline{\alpha}}}(q'))dq'
\end{eqnarray}
with $\bar{H}$ indicating the Hamiltonian for anti-neutrinos. In the case of anti-neutrinos, Eq.(\ref{e:28}) holds but by replacing $H\leftrightarrow\bar{H}$ and  $\nu\leftrightarrow\bar{\nu}$.

Using the results just derived one can identify two interesting conditions to be fullfilled : i)  for the neutrino-neutrino interaction effects to occur; ii) for the onset of bipolar oscillations.
Let us first focus on the  the start of the synchronization regime and discuss the value of $\dot{\tilde{\beta}}$ at $x=R_{\nu}$. In particular, for such a quantity to be non zero, it is necessary $\dot{H}_{e\mu} \neq 0$ and not parallel to $H_{e\mu}$. If  $H_{e\mu}$ is pure real then $\dot{H}_{e\mu}$ has to have an imaginary component. 
Using Eqs.(\ref{e:24}-\ref{e:25}) and (\ref{e:28}) one obtains

\begin{eqnarray}
\left[ H,\rho_{\nu_{\underline{e}}} \right]_{e\mu} &=& -H^{vac}_{e\mu} \\
\left[ H,\rho_{\nu_{\underline{\mu}}} \right]_{e\mu} &=& H^{vac}_{e\mu}
\end{eqnarray}

\noindent
and the same for anti-neutrinos but with $\bar{H}$ instead of $H$. By taking the explicit vacuum terms 

\begin{equation}
H^{vac}_{e\mu}=\bar{H}^{vac}_{e\mu}=\dfrac{\Delta m^2}{4
q'}\sin2\theta
\end{equation}
one gets
\begin{equation}\label{e:51bis}
\int dq'\;\left[ H,\rho_{\nu_{\underline{e}}} \right]_{e\mu}
L_{\nu_{\underline{e}}}
=\dfrac{\Delta m^2}{4}\sin2\theta \int dq' \dfrac{L_{\nu_{\underline{e}}}}{q'}
\end{equation}
If we replace the neutrino number flux at the neutrinosphere by e.g. the Fermi-Dirac distribution for the $\nu_{\underline{e}}$ contribution,  one obtains
\begin{equation}\label{e:51}
\int dq'\;\left[ H,\rho_{\nu_{\underline{e}}} \right]_{e\mu}
L_{\nu_{\underline{e}}}
=-\dfrac{\Delta m^2 L_0}{4\langle E_{\nu_e} \rangle^2}\sin2\theta \dfrac{F_3(\eta)
F_1(\eta)}{F^2_2(\eta)}
\end{equation}
with $F_i(\eta)$ the complete Fermi-Dirac integrals (without the gamma function normalization) and
\begin{equation}\label{e:52}
\dot{H}_{e\mu}=-i\,c\,\sin2\theta  \left(
\dfrac{2}{\langle E_{\nu_\mu} \rangle^2}
-\dfrac{1}{\langle E_{\nu_e} \rangle^2}-\dfrac{1}{\langle
E_{\nu_{\bar{e}}} \rangle^2} \right)
\end{equation}
where $\langle E_{\nu} \rangle$ indicates the neutrino average energies and
\begin{equation}\label{e:53}
c = \dfrac{\sqrt{2}G_F}{4\pi R^2_\nu} \dfrac{\Delta
m^2}{4} L_0 \dfrac{F_3(\eta) F_1(\eta)}{F^2_2(\eta)}
\end{equation}
It has been pointed out in Ref.\cite{Hannestad:2006nj}, in the analogy with a pendulum, and also in Ref.\cite{Fogli:2007bk}
that the vacuum mixing angle need to be non-zero for the neutrino-neutrino interaction to have an effect. 
In our case, where we exactly solve the two-neutrino evolution with all contributions in the 'matter' basis,
since the dominating term comes from $\dot{\tilde{\beta}}$ we find that the condition to have effects coming from the neutrino-neutrino interaction needs Eqs.(\ref{e:51bis}) and (\ref{e:52}) to be non-zero, i.e. if $\sin2\theta \ne 0$, in agreement with  Ref.\cite{Hannestad:2006nj,Fogli:2007bk}.

Let us now focus on the onset of bipolar oscillations.
In Ref.\cite{Duan:2010bf}, within a three flavor multiangle calculation, it has been pointed out heuristically that the transition to the bipolar regime should be due to a condition involving the vacuum and the neutrino-neutrino interaction contributions.
Here with only two flavors a condition signifying the end of the synchronization regime can be explicitly identified from $\dot{\tilde{\beta}}$ Eq.(\ref{e:23}) namely that $|\Hem|^2$ has to approach zero. 
More explicitly, the element $\Hem$ only involves the vacuum and self-interaction terms so we find the onset of the bipolar regime occurs when:
\begin{equation}\label{e:35}
 \left| \Hem \right|^2 =  \left| \Hem^{vac} \right|^2 +  \left|
\Hem^{\nu\nu} \right|^2 + 2\ \mathcal{R}( \Hem^{vac}\Hem^{*\ \nu\nu} )
\rightarrow 0
\end{equation}
This is agreement with the heuristic condition given in \cite{Duan:2010bf}. Note that, although obtained in two flavours, our condition (\ref{e:35}) is obtained from the equations derived in section II.C where no simplifying assumption has been made regarding the neutrino Hamiltonian.  


\subsection{Numerical results and discussion} 
\label{sec:3flavour}
\noindent
Here we present our two-neutrino flavor numerical results in the 'matter' basis obtained using 
Eqs.(\ref{e:18}),(\ref{e:23}),(\ref{e:27}-\ref{e:28}). 

In our calculations the parameters are fixed as follows. For the mixing parameters we take 
 $|\delta m^2|= 2.4 \times 10^{-3}$eV$^2$, $\theta=9^\circ$. 
 For the matter density profile we use 
$\rho_B=1.5 \times 10^8\,(10/x)$ in units of g.cm$^{-3}$ and of km (for $x$).
The neutrinosphere $R_{\nu}$ is taken at 10 km and the corresponding neutrino fluxes are assumed to be of Fermi-Dirac type with average energies  $ \langle E_{\nu_e} \rangle =12$ MeV, $ \langle E_{\bar{\nu}_e}\rangle=15$ MeV and $ \langle E_{\nu_{x}} \rangle=18$ MeV. Equipartition of energy is assumed, with a total luminosity of $4\times10^{51}$ erg.s$^{-1}$, so that in our calculations collective effects only appear in inverted hierarchy. However our findings concerning the role of the matter phase are expected to hold also in the case where equipartition is not satisfied and multiple spectral splits appear.

\begin{figure}[h]
	\vspace*{1mm}
	\centering
		\includegraphics[width=90mm,scale=0.3]{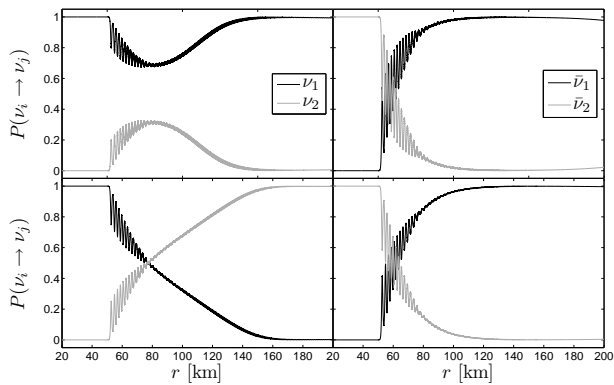}
	\caption{Two-flavor neutrino evolution in the 'matter' basis:
	The figures show the neutrino (left) and anti-neutrino (right figures) survival probabilities for first (solid black) and second (solid grey) matter eigenstates as a function of distance within a core-collapse supernova for energies of $5\;{\rm MeV}$ (upper) and $10\;{\rm MeV}$ (lower) figures.}
	\label{fig:proba}
\end{figure}

\begin{figure}[h]
	\vspace*{1mm}
	\centering
		\includegraphics[width=90mm,scale=0.3]{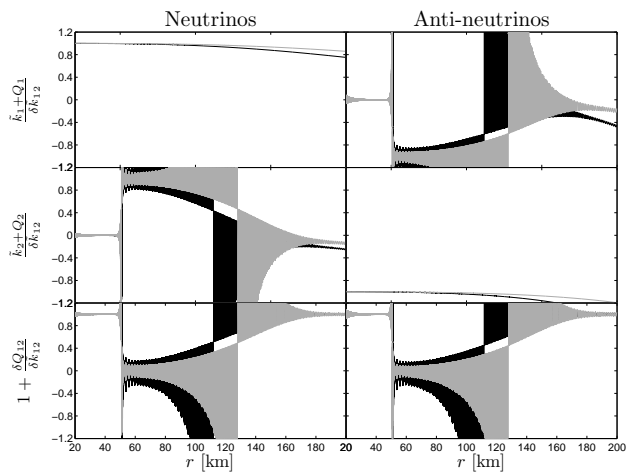}
	\caption{Two-flavors neutrino evolution in the 'matter' basis:
	Diagonal elements of the Hamiltonian Eq.(\ref{e:18}) $(\tilde{k}_1 + Q_{1})/\delta \tilde{k}_{12}$ (upper),
	$(\tilde{k}_2 +  Q_{2})/\delta \tilde{k}_{12}$ (middle) and their difference 
	$1 + \delta Q_{12}/\delta \tilde{k}_{12}$ (lower figures). The curves correspond to
	a 5 (black) and 10 MeV (grey) neutrino energy 
	for neutrinos (left) and anti-neutrinos (right figures).
	The black (grey) lines show the average for a 5 (10) MeV neutrino.
	The calculations include the vacuum mixing, the coupling to matter and the neutrino-neutrino interaction.}
	\label{fig:1}
\end{figure}
\begin{figure}[h]
	\vspace*{1mm}
	\centering
		\includegraphics[width=90mm,scale=0.3]{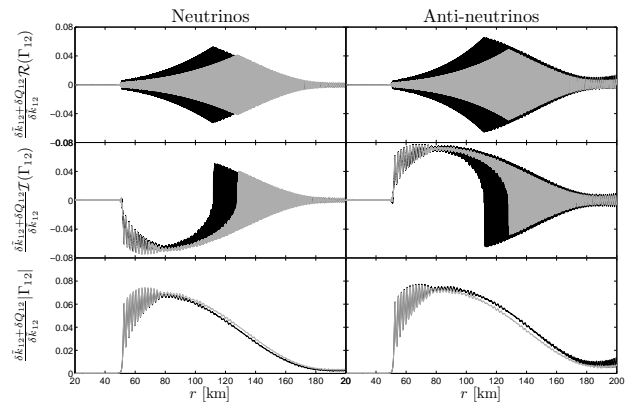}
	\caption{Same as Figure \ref{fig:1} but for the off-diagonal element of 
	two-neutrino flavor Hamiltonian in the basis of the instantenous eigenstates in matter. The results correspond to the real (upper),
	imaginary (middle) parts of $\Gamma_{12}$ and its modulus 
	$|\Gamma_{12}|$ (lower figures), multiplied by  the factor
$(\delta \tilde{k}_{12} + \delta Q_{12})/\delta \tilde{k}_{12}$ .}
	\label{fig:2}
\end{figure}

Figure \ref{fig:proba} shows the oscillation probabilities of the neutrino instantenous eigenstates in matter, as a function of the distance in the supernova for two different neutrino energies. The case considered here is inverted hierarchy.
The synchronization regime, bipolar oscillations and the spectral split are
easily recognized. The first corresponds to the region where neutrino flavour conversion is frozen; while in the second regime oscillations appear in the probabilities.
The spectral split phenomenon occurs between 70 and 100 km and correspond to
the complete flavour conversion, giving rise to a swap of the neutrino spectra, 
depending on the neutrino energies. As an example,
the two energies shown are smaller (larger) than the spectral critical energy for which no (full) neutrino flavour conversion takes place.

Figure \ref{fig:1} presents the diagonal $\left(\Tk{i}+Q_{i}\right)/\dTk{12} $ with $i=1,2$ calculating using Eqs.(\ref{e:theta}) and (\ref{e:23}), as well as the difference of the diagonal elements $1+\dQ{12}/\dTk{12} $. It is indeed this quantity which is important to follow the neutrino flavor evolution. The results are shown both for neutrinos and anti-neutrinos and different neutrino energies. One can see that, since $\tilde{\theta}$ is minimum ($ \approx 0$) for neutrinos and maximum ($\approx \pi/2$) for anti-neutrinos, one of the diagonal matrix elements is approximately given by $\tilde{k}_i$ ($\tilde{k}_1$ for neutrinos and $\tilde{k}_2$ for anti-neutrinos) while the other oscillates very fast due to $\dot{\tilde{\beta}}$. 

\begin{figure}[h]
	\vspace*{1mm}
	\centering
		\includegraphics[width=90mm,scale=0.3]{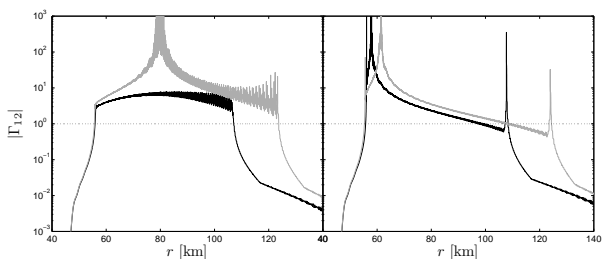}
	\caption{Ratio of the average of the modulus of the numerator over the average of the denominator of the generalized adiabaticity parameter $\Gamma_{12}$ Eq.(\ref{e:Gamma12}) corresponding to the 
	two-neutrino flavor Hamiltonian in the 'matter' basis, for neutrinos (left) and for anti-neutrinos (right figure), with an energy of
5 (black) and 10 MeV (grey).}
	\label{fig:2b}
\end{figure}

The real and imaginary part of the off-diagonal contributions 
$\Gamma_{12}(\delta \tilde{k}_{12} + \delta Q_{12})/\delta \tilde{k}_{12}$ Eq.(\ref{e:Gamma12})    of the Hamiltonian in the 'matter' basis Eq.(\ref{e:18}) are presented in Figure \ref{fig:2}. Results are given for both neutrinos and anti-neutrinos of different energies. Comparing the scales of Figures \ref{fig:1} and \ref{fig:2} one sees that the diagonal elements are much larger than the off-diagonal ones as expected.
The off-diagonal contributions are practically zero in the synchronization region below 50 km where they abruptly become non-zero.
Fast oscillations coming from the derivative of the phase are again present. Note that the numerical results are very similar both for
neutrinos and anti-neutrinos. 
Figure \ref{fig:2b} presents the ratio of the average of the modulus of the numerator over the average of the denominator of the generalized adiabaticity parameter $\Gamma_{12}$ Eq.(\ref{e:Gamma12}).
One can see that such a quantity is larger than one in the region where the
neutrino-neutrino interaction effects are dominant.

Let us now discuss the relative role of the diagonal and off-diagonal terms of the Hamiltonian in the 'matter' basis, when the neutrino-neutrino contribution is included.
One first notes that,
in absence of such a contribution, $\dQ{12}=0$ and $\delta \tilde{k}_{12}$ would not vary over the region of 200 km near the neutrinosphere. However,
after the inclusion of $H_{\nu\nu}$, the difference between the diagonal entries in $\tilde{H}$, i.e. $\dTk{12}+\dQ{12}$, drops suddenly at around 50 km when the bipolar oscillations begin (bottom part of Figure \ref{fig:1}). This is due to the fact that 
one of the two diagonal elements suddenly changes due to the $\dot{\Tb}\cos^{2}\Tt{}$  ($\dot{\Tb}\sin^{2}\Tt{}$) for neutrinos (anti-neutrinos). 
As a consequence, when bipolar oscillations start, the difference 
between the diagonal elements oscillates around zero, due to the difference $\dQ{12}=\dot{\tilde{\beta}}\cos2\tilde{\theta}$. These results 
make it clear that the Dirac phase $\Tb$ acquires a particularly significant role in the neutrino flavor evolution. It leads to an oscillatory degeneracy between the diagonal elements of $\tilde{H}$, increasing significantly the importance of the off-diagonal entries in $\tilde{H}$ which 
had a negligible role during the synchronization regime.
Indeed, in the region where the neutrino-neutrino interaction effects are dominant the derivative of the matter mixing angle is $\dot{\tilde{\theta}} \approx 0$. The size of the off-diagonal elements of $\tilde{H}$ is approximately $\dot{\tilde{\beta}}\tilde{\theta}$. 
It is clear that these small off-diagonal contributions become very important when the difference of the diagonal elements falls to zero.

\begin{figure}[h]
	\vspace*{1mm}
	\centering
		\includegraphics[width=90mm,scale=0.3]{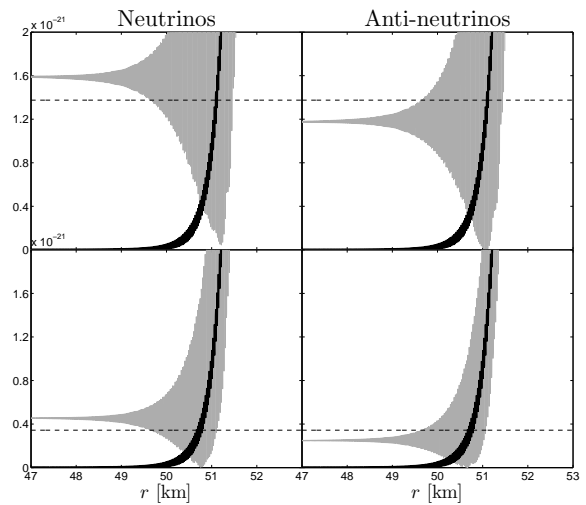}
	\caption{Contributions to the off-diagonal matrix elements  $|H_{e\mu}|$ Eq.(\ref{e:35}) of the two-neutrino flavor Hamiltonian. The curves show : $|H^{vac}_{e\mu}|^2$ (dashed), $|H^{\nu\nu}_{e\mu}|^2$ (black) and the total $|H_{e\mu}|^2$
	(grey), in units of $eV^2$, for neutrinos (left) and anti-neutrinos (right). Results for a 5 MeV neutrino are given in the upper figures and for a 10 MeV one in the lower figures.}
	\label{fig:3}
\end{figure}


Concerning the onset of bipolar oscillations,
Figure \ref{fig:3} 
shows that the different terms corresponding to the condition $|H_{e\mu}|^2=0$ Eq.(\ref{e:35}).
At the beginning $H_{e\mu}$ is dominated by the constant vacuum contribution.      
Since the modulus of $|H^{\nu\nu}_{e\mu}|^2$ increases as the interference between these two terms. 
Only when these contributions have been cancelled off can the phase $\Tb$ vary rapidly.
So the point where $|H^{vac}_{e\mu}|=|H^{\nu\nu}_{e\mu}|$ engenders an abrupt change in $Q_{1}$ and $Q_{2}$ (Figure \ref{fig:1}) and leads to a concomitant sudden increase in $\Gamma_{12}$.
When the neutrino-neutrino interaction term becomes negligible another change in the $Q_{i}$ occurs. From Figure \ref{fig:1} this appears to occur at an energy dependent distance of 110 km for $5\;{\rm MeV}$ and 130 km for $10\;{\rm MeV}$ neutrinos, where the neutrino probabilities have reached their asymptotic behavior.

\section{Conclusions} \label{sec:conclusions}
General expressions for the diagonal and off-diagonal  terms of the Hamiltonian 
 in the basis of the instantenous eigenstates in matter have been derived,
corresponding to a general Hamiltonian in the flavour basis with complex contributions.  Relations have been given to determine in particular the off-diagonal contributions as a function
of the neutrino mixing matrix relating the flavor basis and the one formed by instantenous eigenstates in matter. We have applied these findings to the case of
neutrino propagating in a core-collapse supernova in the two-neutrino case. Our analytical results show that the flavor evolution is governed by the derivatives of both the mixing angle and a phase. The numerical results show the important role of the matter phase. In particular, 
the diagonal and off-diagonal contributions to the Hamiltonian, in the 'matter'basis, show abrupt changes at the end of the synchronization region that we have associated to a divergence condition for the derivative of the matter phase, involving the vacuum and the matter term. Such a condition, that characterizes the onset of bipolar oscillations, is only slightly dependent on the neutrino energy. In future work we plan to extend the present study to the case of three flavors, in order to determine whether additional insight may be gathered.

\section{Appendix A}
\noindent
Here we show that the Majorana phases in matter do not influence the neutrino flavour conversion. In fact, there exists a basis in which the contributions to the diagonal matrix elements of the matter Hamiltonian are absent. This occurs if we introduce the adiabatic basis, discussed in \cite{Kneller:2009vd}, related to the matter basis via the unitary transformation $W(x)$ i.e. $\tilde{\psi} = W\,\psi^{(a)}$. In this new basis, the Schr\"odinger equation is 
\begin{subequations}
\begin{eqnarray}
i \frac{d\psi^{(a)}}{dt} & = & \left( W^{\dag}\Tilde{K}W -i W^{\dag}\frac{dW}{dt} -i W^{\dag}\TU^{\dag}\frac{d\TU}{dt}W
\right)\,\psi^{(a)}, \nonumber \\ 
\\ & \equiv & H^{(a)}\psi^{(a)}.
\end{eqnarray}
\end{subequations}
The matrix $W$ is chosen so that it removes the diagonal elements of $H^{(a)}$ including the derivatives of the Majorana phases that appear in the diagonal entries of $\TU^{\dag}\,d\TU/dt$. This requirement indicates that $W$ is diagonal so writing $W$ as
\begin{equation}
W=\left(\begin{array}{ccc} \exp[-2\,i \pi \phi_{1}] & 0 & \ldots \\ 0 &
\exp[-2\,i \pi \phi_{2}] & \ldots \\ \vdots & \vdots & \ddots
\end{array}\right) 
\label{eq:W}
\end{equation}
one obtains that the phases $\phi_{i}$ are defined by 
\begin{equation}
\frac{d\phi_{i}}{dt} = \frac{1}{2\,\pi}\left( \Tk{i} + Q_{i} -\frac{d\alpha_{i}}{dt} \right). \label{eq:dphidx}
\end{equation}
Integrating gives
\begin{equation}
\phi_{i}(t) = \frac{1}{2\,\pi}\left[ - \Ta{i}(t) + \int dt' \left( \Tk{i} + Q_{i} \right) \right]. \label{eq:phi}
\end{equation}
Thus we find that $W$ factors into a matrix $\slashed{W}$ independent of the $\Tilde{\alpha}$'s, and the Majorana matrix $\TA$, i.e. $W=\TA^{\dagger}\,\slashed{W}=\slashed{W}\,\TA^{\dagger}$. 
When we insert this solution for $W$ into $H^{(a)}$ we find that the Majorana phase dependence in the off-diagonal entries of $H^{(a)}$ disappears too i.e.  
\begin{equation}
 H^{(a)} = -iW^{\dag}\TU^{\dag}\frac{d\TU}{dt}W = -i \slashed{W}^{\dag}\slashed{\TU}^{\dag}\frac{d\slashed{\TU}}{dt}\slashed{W}.
\end{equation}
Thus we are able to find a basis in which the Hamiltonian is entirely independent of the Majorana phases which, in turn, implies the S-matrix $S^{(a)}$ is also independent of 
the phases in this basis. The difference between matter basis and this new basis is simply a position dependent rephasing of the elements in the Hamiltonians and the associated S matrices.  Thus we reach the expected conclusion that that the Majorana phases we introduced in the matter mixing matrix cannot affect neutrino oscillation observables regardless of what we pick for them. As a consequence, in our derivations, we can set the Majorana phases to constants (or zero) and neglect them (as well as their derivatives), without any loss of generality.

\section{Appendix B}
\noindent
We furnish here expressions for the derivative of the neutrino-neutrino interaction Hamiltonian Eq.(\ref{e:1tris}), that can be of use in future multi-angle calculations of neutrino evolution in a core-collapse supernova, using, as in the present manuscript, the 'matter' basis. Assuming spherical geometry for the neutrinosphere, the non-linear Hamiltonian Eq.(\ref{e:1tris}) reads
\begin{eqnarray}\label{e:29}
H_{\nu\nu}^{m.a.}&=&\dfrac{\sqrt{2} G_F}{2 \pi R_{\nu}^2}
 \sum_{\alpha} \int^{\infty}_{0} \int^{1}_{\cos\theta_{max}} dq'\,d\cos\theta'\left(1-\cos\theta \cos\theta'\right) \nonumber\\
&&[\rho_{{\nu}_{\underline{\alpha}}} (q',\theta') L_{{\nu}_{\underline{\alpha}}}
(q') - \rho_{\bar{{\nu}}_{\underline{\alpha}}}^*(q',\theta') L_{\bar{{\nu}}_{\underline{\alpha}}}(q')]
\end{eqnarray}
with:
\begin{equation}
 \cos\theta_{max}=\sqrt{1-\left(R_\nu/x\right)^2}
\end{equation}
Using the Leibniz Integral Rule :
\begin{eqnarray}\label{e:30}
\dfrac{\partial}{\partial z} \int^{b(z)}_{a(z)} f(y,z) dt &=& \int^{b(z)}_{a(z)} \dfrac{\partial f(y,z)}{\partial z} dt \nonumber\\
&&+ f(b(z),z) \dfrac{\partial b(z)}{\partial z} \nonumber\\
&&- f(a(z),z) \dfrac{\partial a(z)}{\partial z}
\end{eqnarray}
The derivative for Eq.(\ref{e:29}) is 
\begin{equation}
 \dot{H}_{\nu\nu}^{m.a.} = \dfrac{\sqrt{2} G_F}{2 \pi R_{\nu}^2} \sum_{\alpha} \int^{\infty}_{0}dq'\,
 \left( R(\theta',q')- S(x)\,T(q') \right)
\end{equation}
with
\begin{subequations}
 \begin{eqnarray}\label{e:31}
R(\theta',q')&=& \int^{1}_{\cos\theta_{max}}d\cos\theta'\,\left(-i \left(1-\cos\theta \cos\theta'\right)\right. \nonumber\\
&&\left(\left[ H, \rho_{{\nu}_{\underline{\alpha}}}(q',\theta')\right] L_{{\nu}_{\underline{\alpha}}}(q') \right. \nonumber\\
&&\left.\left. + \left[ \bar{H}, \rho_{{\bar{\nu}}_{\underline{\alpha}}}(q',\theta')\right]^* L_{\bar{{\nu}}_{\underline{\alpha}}}(q')\right)\right) \\
S(x)&=& \dfrac{\left(R_\nu /x\right)^2}{x\sqrt{1-\left(R_\nu/x\right)^2}}\\
T(q')&=& \left(1-\cos\theta \cos\theta_{max}\right)[\rho_{\nu_{\underline{\alpha}}} (q',\theta_{max}) L_{\nu_{\underline{\alpha}}}(q') \nonumber\\
&& - \rho_{\bar{{\nu}}_{\underline{\alpha}}}^*(q',\theta_{max}) L_{\bar{{\nu}}_{\underline{\alpha}}}(q')]
\end{eqnarray} 
\end{subequations}

\newpage


\end{document}